REVIEW

# Making Sense of Computational Psychiatry

Lilianne R. Mujica-Parodi, Helmut H. Strey

Department of Biomedical Engineering (Drs Mujica-Parodi and Strey), and Laufer Center for Physical and Quantitative Biology (Drs Mujica-Parodi and Strey), Stony Brook University, Stony Brook, New York.

Correspondence: Lilianne R. Mujica-Parodi, PhD, Director, Laboratory for Computational Neurodiagnostics, Professor, Department of Biomedical Engineering, Renaissance School of Medicine, Stony Brook, NY 11794-5281 (Lilianne.Strey@stonybrook.edu) or Helmut H. Strey, PhD, Department of Biomedical Engineering, Stony Brook University, Stony Brook, NY 11794-5281 (Helmut.Strey@stonybrook.edu).

## Abstract

In psychiatry we often speak of constructing "models." Here we try to make sense of what such a claim might mean, starting with the most fundamental question: "What is (and isn't) a model?" We then discuss, in a concrete measurable sense, what it means for a model to be useful. In so doing, we first identify the added value that a computational model can provide in the context of accuracy and power. We then present limitations of standard statistical methods and provide suggestions for how we can expand the explanatory power of our analyses by reconceptualizing statistical models as dynamical systems. Finally, we address the problem of model building—suggesting ways in which computational psychiatry can escape the potential for cognitive biases imposed by classical hypothesis-driven research, exploiting deep systems-level information contained within neuroimaging data to advance our understanding of psychiatric neuroscience.

## Introduction

Psychiatry frequently refers to models. At its very basis, a model is a heuristic, a way to make sense of a complex set of interactions and relationships by virtue of a simple rule. Psychiatry's earliest models attempted to explain the psyche, and therefore deviations from its norms, using heuristics based on psychoanalytic theories on the influence of, often unconscious, childhood experiences and defense mechanisms (Figure 1A). These conceptual models persist today (e.g., the "attachment parenting" model identifies various behavioral pathologies emerging later in life as the consequence of inadequate parental responsiveness during a child's early years (Benoit, 2004). However, as psychiatry has embraced rapid gains made possible by noninvasive neuroimaging, most current psychiatric models now explicitly incorporate information about the brain, with corresponding interest in "circuits." These tend to integrate neural and psychological frameworks; for example, by describing "regulation" within region-of-interest–scale systems schematically described by arrows connecting boxes representing regions associated with different psychologically defined functions (e.g., "fear," "craving," and "willpower").

In contrast, the term "circuit" in basic neuroscience is typically used to refer to microcircuits in which biophysical processes, such as changing resting membrane potentials, modulate signal response. As with recent elegant work on thalamocortical adaptive sensory gating (Mease et al., 2014; Manita et al., 2015), these can take the form of complex control processes, which are normally derived from rodent electrophysiology, using optogenetic and/or chemogenetic manipulations. While some clinical neuroscience models (Figure 1B) are also dynamical systems as defined by control systems engineering (Figure 1C) (i.e., they describe systems as a whole that can predict trajectories over time), the more typical use of the term "circuit," as currently used within the clinical neuroscience literature, reflects co-activation between regions, defined by correlations. These co-activated structures are actually more appropriately described as "networks" than circuits, a distinction that contributes to the conceptual

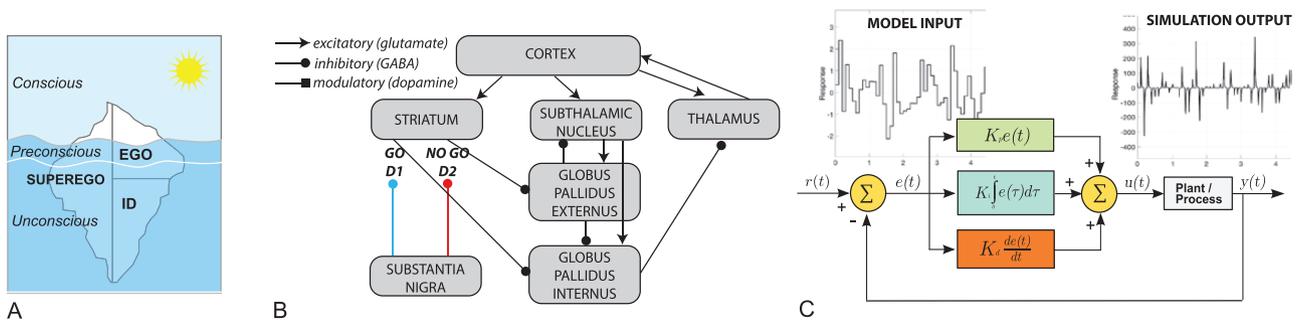

**Figure 1.** Three "models": Freudian iceberg model of the human mind, modern neuropsychiatric control circuit model, and dynamical systems model of a control circuit. (A) An example of the typical conceptual models historically grounding psychiatry, Freud's theories assume that behavior reflects unconscious influences provided by components of the mind, including drives ("id"). (B) Schematic of the *cortico-basal ganglia-thalamo-cortical* (CBGTC) loop, integrating data from multiple neuroimaging (MR-PET) modalities. Crucially, the CBGTC loop can be treated either as a heuristic (as per the Freudian iceberg model) in which it functions more as a map than a true model—meaning that it cannot be operationalized to make predictions over time—with new inputs, or it can be treated as a graphical representation of a dynamical system of differential equations as per (Frank, 2005) and the PID control circuit shown in C. (C) The dynamical system model is distinguished by the fact that, when presented with time-series inputs, it produces simulation outputs. These output trajectories can be compared with new data to rigorously assess the model's validity. Images adapted from (Maia and Frank, 2011) and Arturo Urquizo (CC-BY 3.0).

divide between clinical neuroscience, basic neuroscience, and their computational extensions.

The integration of human neuroimaging with computational neuroscience is comparatively recent, with the field today broadly consisting of 3 approaches that all fall within the description of dynamical systems. The first are bottom-up biophysical approaches that start with neurons and then extrapolate to aggregates and their mean field dynamics, for example, Magneto/electroencephalography (M/EEG) oscillations (Schurmann et al., 2007; Murray et al., 2018). The second are top-down approaches, which start from emergent phenomena and try to infer a set of neural mechanisms that could lead to such phenomena (Deco et al., 2011; Radulescu and Mujica-Parodi, 2014; Breakspear, 2017; Pillai and Jirsa, 2017). The third, information-theoretic, approach investigates structural strategies by which the brain might optimize for efficiency in information propagation based on considerations from graph theory/network science (Bullmore and Sporns, 2009; Bassett et al., 2010; Gu et al., 2015).

Excellent reviews can be found in the cited references for each of the above topics. Therefore, our scope here will be more general and attempt to address 2 fundamental questions. First, how does mainstream neuropsychiatry differ in its models as compared with more mature fields like physics, the latter of which are closer to control/dynamical systems as employed by computational psychiatry? Second, what follows from these differences? In so doing, we hope to "make sense" of computational psychiatry and why we believe it offers the potential for transformative advances for the neuropsychiatric field, both scientific and clinical.

## The Danger of Category Errors

Psychiatric/psychological models have historically distinguished themselves from those in neuroscience by their focus on narrative (e.g., the "schizophrenogenic mother" model, in which early maternal rejection causes her child's later psychotic paranoia) (Morris, 2012). At its best, this narrative approach has allowed psychiatry to move beyond mere taxonomies of signs and symptoms to approach the mind in a manner that aims to be mechanistic and causal. At its worst, it risks providing a satisfyingly intuitive, yet ultimately hollow, "just so" story with no manner of objectively confirming or disconfirming the correctness of its conceptual frameworks. Unfortunately, the effects associated with unconfirmed conceptual frameworks are not localized to the specific interpretation of results that promote them but can be more generally pernicious if they inform the generation and interpretation of new data. To the degree to which a field permits untested models to propagate, it risks following false leads that can take generations of experiments to correct.

One recent example is the psychology field's model of *oxytocin*—a neuropeptide endogenously produced during female labor and lactation. The conceptual model of oxytocin as a "trust drug" was based on the assumption that hormonal influences on maternal bonding might extend to blind induction of "trust." Behavioral responses (greater generosity in the face of nonreciprocal feedback) (Kosfeld et al., 2005) and neurobiological responses (weaker amygdala activation) (Baumgartner et al., 2008) to exogenous administration of oxytocin when patients played several variations of a neuroeconomic game, as well as subsequent neuroimaging experiments of fear (Kirsch et al., 2005), conditioned fear (Petrovic et al., 2008), and fearful faces (Domes et al., 2007; Gamer et al., 2010), seemed initially to support this model. However, after these results were embraced by the media and the clinical community, which went so far as to propose clinical trials to test the administration of oxytocin in autistic children, the oxytocin as a pro-social drug of love and trust model had to be radically reconsidered in the face of contradictory data. Oxytocin does sometimes appear to increase trust and altruism, but it also sometimes appears to decrease them (Shamay-Tsoory et al., 2009; De Dreu et al., 2010, 2011; Bartz et al., 2011; Grillon et al., 2013; Ne'eman et al., 2016); indeed, many of the same results used to justify the "trust" role are equally consistent with the attenuation of reinforcement learning (Ide et al., 2018), an alternative "model" that has nothing to do with "trust" per se.

On the one hand, it is important to emphasize that the paradigm shift in understanding oxytocin does not represent a failure of science; the fact that the field ultimately recognized a contradiction between its model and subsequent data (albeit after approximately 1200 published articles) is a measure of its success. But the rise and fall of oxytocin as "trust drug" illustrates our dependence on models and our need to create them responsibly, because a flawed conceptual framework (e.g., oxytocin as "trust drug" or, far more perniciously, the "schizophrenogenic mother") persisted for so long precisely because it circularly restricted the kinds of hypotheses the field formulated in testing it. In logic, this type of dependence on false assumptions is known as a "category error." If I ask a bachelor "Did you murder your wife this morning or last night?" I am asking the wrong type of question, since the bachelor never had a wife and therefore could not have murdered her—at either (or indeed any) time. If I run statistical tests on results obtained from a large group of bachelors and formulate "Murdered your wife this morning" as the test hypothesis and "Murdered your wife last night" as the null hypothesis, I am still asking the wrong type of question, and no amount of additional data or multivariate analyses will get us any closer to the truth.

In an effort to avoid the "fishing expeditions" that lead to statistical errors of multiple comparisons (i.e., in which the probability of error associated with *P* values for each test multiplies with the total number of tests conducted), modern psychiatry currently holds "hypothesis-driven" research as a gold standard, since it minimizes the number of questions asked and therefore tests conducted. Yet, the system being measured obviously does not know or care how many statistical tests are being conducted on its data, and restricting oneself to asking a specific question when approaching a system about which we know very little greatly increases the inherent probability that we are asking the wrong question, that is, committing a category error. It is as if we are trying to find a well-hidden treasure, and instead of asking its keeper, a genie, "Genie, where is the treasure?" we ask "Genie, is the treasure in the bedroom? The kitchen? The garden?" If, in fact, our model of the treasure being in the house is incorrect—rather, the treasure is in a different country altogether (or under the sea)—it will take us a very long time to guess the right question to ask. Thus, when approaching a system about which we know very little, it can be more strategic, efficient, and honest to start by eschewing model-specific biases, because models based on false assumptions run the risk of leading us (and the field) astray.

A central argument that we will try to make here is that nearly all of psychiatry and psychiatric neuroscience is a system about which we still know very little, and thus its models are particularly vulnerable to category errors. The source of the first, more obvious, vulnerability is the inherent difficulty in precisely and objectively defining its terms: measuring an individual's "attachment" to a parent, or "willpower" in resisting temptation, or "trust" in a stranger, or the severity of his depression or psychosis cannot be done directly, but only in ways that make multiple (and often themselves ill-defined) secondary assumptions within the model, a problem only compounded by the measures' dependence on self-report. This is in contrast to most measurements made in classical physics; for example, determining the temperature of a glass of water uses a measurement device (thermometer), whose relationship between inputs and outputs (temperature, movement of mercury) is both precise and mechanistic. The second vulnerability, on which we will want to focus in this article, is perhaps less obvious: the psychiatric field's historical dependence on hypothesis-driven statistical analyses regularly assumes models but has largely neglected to develop methods for rigorously constructing, testing, and validating them. Below, we will first address how models are used in other fields (e.g., physics), and why the common standard of performing statistical tests and evaluating the robustness of its results according to some standard threshold within the field (e.g., *P* values) is a process that is not designed to rigorously construct, test, or validate the kind of models most relevant to psychiatry. We then follow by suggesting how computational modeling of dynamical systems—particularly its Bayesian and generative extensions—may provide added value to psychiatry by providing directional guidance with both building and validating its conceptual frameworks in a manner that avoids statistical problems of multiple comparison but also avoids premature bias to models before they are validated. To avoid conflating the first and second vulnerabilities, we will primarily use examples from functional neuroimaging, which ostensibly provide greater assurance that we know what we are measuring (here assumed to be blood oxygenation level dependent (BOLD) activation, produced via neurovascular coupling from neuronal response, with all the usual caveats regarding the reliability of neurovascular coupling and the imperfect signal/noise of functional magnetic resonance imaging [fMRI]). However, the same principles can be extended to any modality in psychiatry, including behavioral models.

## Testing and Validating Models

Operationally, models have a unique and extraordinarily useful feature. By including both structural (cause *c* is connected to effect *e*) and functional (the behavior of *e* changes in specific ways as a function of *c*) information, they act not only to imbue cognitive understanding of the system but to act as an input-output device: providing a mechanism by which one can provide a system with new input(s) and use the model to predict the system's response (as per the electrical circuit model in Figure 1C). A classic illustration from Newtonian mechanics is the billiard table with a set of known initial conditions, which include some particular configuration of balls in space at time $t_0$. If one hits a ball from some known direction with some known force, our model (e.g., the combined equations that govern the laws of motion) will predict a changed configuration of balls in space at some later time $t_1$. Likewise, in a linear regression: $y = b_0 + b_1x_1 + b_2x_n\dots + \varepsilon$, we could model the influence of genotype and various environmental factors ($x_1$, $x_2$, …) by fitting the optimal set of parameters ($b_0$, $b_1$, …, $b_n$) and measurement error $\varepsilon$, to predict a 10-year-old child's depression at the age of 40. Modeling the linear regression on one set of individuals, we can then take a new child, insert his values for $x_i$, and then test whether our model predicts the new child's psychological state 30 years later, *y*.

The utility of a heuristic can be quantified by the degree to which it maximizes the amount of data it predicts while minimizing the amount of data it does not predict. As intuitively illustrated by the billiard ball and depression examples, a model's accuracy can be objectively and quantitatively assessed by the distance between the predicted: that is, the predicted configuration of balls or adult behavior at *t(i)*, vs the actual *y*: that is, the actual configuration of balls or adult behavior at *t(i)*. The model's power can be assessed with respect to the diversity of inputs (e.g., strikes from different angles and with different degrees of force) as well as the length of time *i* into the future over which the model still continues to hold (e.g., accurately predicting the billiard ball configuration 10 seconds post-strike vs only 1 second post-strike, or predicting clinical trajectories decades, rather than weeks, into the future). To provide a direct comparison of these psychiatric models with their counterparts in physics, psychiatric models would first need to articulate a precise set of initial conditions (in this case, objective and measurable features for a brain or mind's "state"). These are the biomarkers ubiquitous to neuropsychiatry. But second, and most critically, they would need to predict with some degree of accuracy how the brain or mind would behave under a new set of initial conditions as well as how it would evolve from that state over time. Thus, validating the utility (i.e., the accuracy and power) of psychiatric models requires that those models be fundamentally prospective, rather than retrospective, providing empirically verifiable predictions, rather than descriptions or explanations, for their narratives. In logic, this causal validation is described formally by "modus ponens": the truth value of *if c→e* (read as: *if c then e*) can be tested by measuring that *c*, therefore *e* while the truth value of *c→e* cannot be tested by measuring *e*, therefore *c*. The reason for this is that the same effect, *e*, can occur with different, mutually exclusive causes: one cannot infer $c_i$ from *e* if it is also the case that: $c_2 → e$. Unlike in physics, prospective model validation in psychiatry is vanishingly rare. In some cases, one can understand that doing so may be difficult: longitudinal studies following youngsters for decades to assess the validity of our models (inferring effects from causes) is a project far more logistically demanding than taking a cohort of adults and working backwards (inferring causes from effects). However, in testing our neuroimaging models, predictive validation is more tractable, as the same circuit can be feasibly probed with multiple inputs with trajectories that evolve over reasonably short periods of time. Thus, if we define a reward circuit on one group of individuals with one particular set of inputs, that circuit—in an independent group of individuals—should be able to predict neural dynamics over time and in response to a new set of inputs (i.e., not simply the inputs used in estimating the circuits). This feature displays the true power of computational psychiatry, since models that remain robust to a diversity of inputs would permit parallel in silica testing of different pharmacological treatment strategies.

In summary, biomarkers—defined as features reliably associated with some specific state—are valuable, because without them we fall prey to the first vulnerability described above (ill-defined variables) and cannot describe our initial state. But those features are not models unless they can tell us something about trajectories that extend beyond the data they were originally designed to fit. To do so, they can only be tested—and thus validated—with prospective, rather than retrospective, study designs.

## Limitations of Statistics in Constructing Psychiatric Models

As psychiatric models evolved away from the narrative (fundamentally conceptual, explicitly causal) style favored by the psychoanalytic tradition in favor of defining neurobiological features specific to populations or conditions, the field began to rely increasingly on statistics to provide data-driven feedback on whether those features were likely to be correct. Models in physics (e.g., $F = ma$) generally behave with sufficient homogeneity that variance in prediction—given the same set of initial conditions—can be attributed solely to measurement error. In contrast, neurobiological biomarkers generally describe their data with fits so unreliable that the model's degree of unreliability also needs to be quantified. This measure of "how likely is it that I'm wrong in identifying a feature" will generally be far too low a standard for models capable of predictive validation, particularly within realistically complex systems (i.e., not simply pairs of variables) that require multiple interactions. To illustrate why, let us assume our previous example of the straightforward linear regression ($y = b_0 + b_1 x_1 + b_n x_n … + \varepsilon$), which—to make things simple—initially will include only 2 variables: *x* and *y*. Thus, increases in *x* (e.g., activation of the amygdala) are associated with decreases in *y* (e.g., deactivation of the ventromedial prefrontal cortex). A fairly typical threshold for accepting the reliability of this relationship might be $P < .001$, which means that, for a representative fMRI sample size of n = 50, and $r = 0.45$, our threshold for reliability has been reassuringly met. Yet a $r = 0.45$ explains only 20% of the data's variance, which means that fully 80% of the data's variance has not been fit by the model!

Now let us consider that *x* and *y* are almost certainly not the only 2 protagonists in this system: for in fact, these 2 regions are just 2 components of a longer chain of interactions, $x ⟷ y ⟷ z …$. Yet, if every step of the model contains 80% (or even 5%) error, those errors quickly propagate throughout the system with every additional interaction. While a typical response to increasing statistical reliability of our models might be to increase sample sizes, in this case, increasing sample sizes simply makes the problem worse. That is because, as sample size increases, even less accurate models achieve the threshold of statistical significance. For example, for n = 600, to meet our threshold of $P < .001$ only $r = 0.16$ is required, thus explaining only 3% of the data's variance and therefore not explaining 97% of it.

It is important to emphasize that this degree of unreliability is occurring already at the level of model estimation (based upon fitting a model to existing data) and not even at the level of model validation (using the model to predict future outcomes). It is as if we were trying to construct a model of Newtonian mechanics based on an observed game of billiards in which our "model" for the relationship between each pair of balls was a linear regression characterized by $r = 0.87$ (for neurobiological data, an effect size so large as to be almost unheard of), explaining 75% of the variance with respect to each interaction between balls. Yet given that a single strike of the cue will trigger perhaps a dozen subsequent interactions between balls, by the time the laws' error propagates between all dozen interactions, prediction would be so unreliable (accurate only $0.75^{12} = 3.2\%$ of the time and thus inaccurate 96.8% of the time) as to make the model completely useless. For modeling systems of interest to neuropsychiatry, the situation is actually much more dire. The brain propagates signals not simply as a linear chain of interactions but rather within circuits that include regulation via feedback. Depending on the type of feedback, such circuits may amplify or reduce input noise in a nonlinear fashion, as has been shown experimentally in genetic circuits (Murphy et al., 2010; Alon, 2019).

In summary, the *P* values ubiquitous to neuropsychiatry are designed to describe our level of confidence with respect to the existence of relationships and differences between anatomical regions, conditions, or population. As such, they give us reason to believe that there are relationships that might be worth modeling but are inadequate in providing the model itself, particularly for the neurobiological control systems (e.g., cortico-basal ganglia-thalamo-cortical [Maia and Frank, 2011], prefrontal-limbic [Mujica-Parodi et al., 2017] circuits) most relevant to neuropsychiatry. That is because linear regressions are only able to predict outputs for systems with a specific topology: one or more parallel direct inputs (Figure 2A). In contrast, most neurobiological responses also involve series (chain) signal pathways (Figure 2B) as well as thresholds/filters, saturation, feedback processes (Figure 2C) etc. that are fundamentally nonlinear—systems that if modeled as multiple linear statistical relationships would quickly propagate errors to levels that would all but eliminate predictive power and therefore the models' validation. As we shall see below, dynamical systems solve the problem of system-based model validation of neurobiologically relevant structures and thus motivate an alternative approach made possible by computational psychiatry.

## Fitting and Validation of Dynamical Systems Models in a Fuzzy World

We ended the last section by noting that, to capture even the most basic features of brain circuits, it is necessary to accommodate topologies fundamentally outside the scope of what can be modeled with linear regression. Thus, each of our

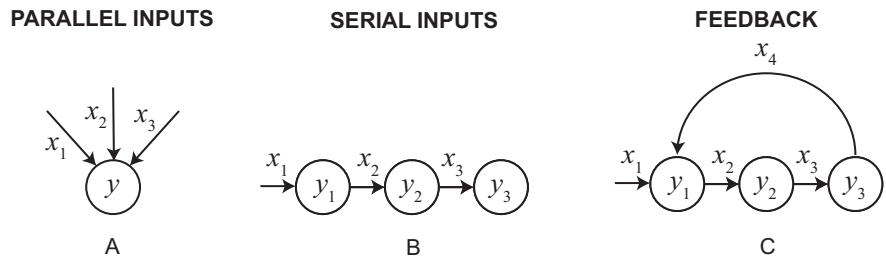

**PARALLEL INPUTS**       **SERIAL INPUTS**       **FEEDBACK**

A                         B                       C

**Figure 2.** Linear regressions are only capable of modeling neurobiological models with a very specific topology: parallel inputs. In contrast to parallel inputs (A), most neurobiological circuits of relevance to psychiatry also require serial (B) and feedback (B) components, structures that could lead to an explosion of error propagation using standard statistical methods. Stochastic dynamical systems, which operate as a system of coupled differential equations, are able to capture all neurobiological-relevant circuit topologies while also providing confidence intervals for not only each individual parameter but also the model as a whole.

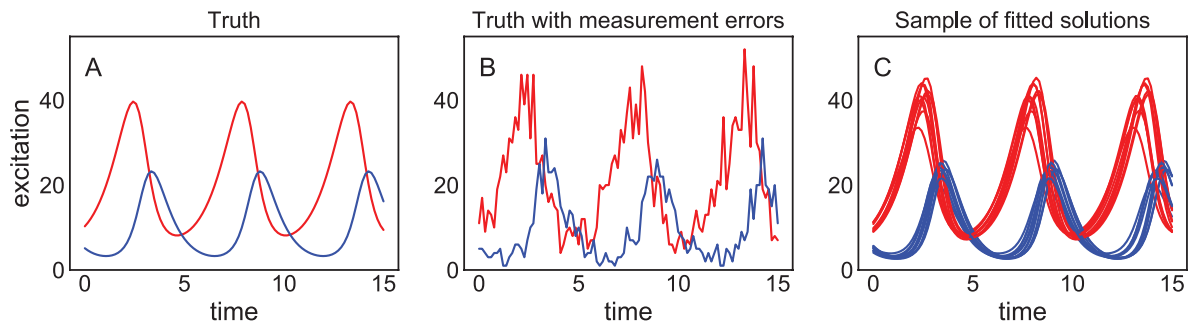

**Figure 3.** Example of how to fit and validate a model in the face of measurement noise. Here we present a simple deterministic dynamical system, Lotka-Volterra (LV), that can be used to describe interplay between excitatory and inhibitory components of a neural circuit with negative feedback. (A) Time course of a specific solution of the LV differential equations (red = excitatory, blue = inhibitory activation). (B) Time course of (A) with added Poisson noise to simulate a realistic data set. (C) Using B as data, we estimated the model parameters of the LV model and plotted a sample of the possible solutions given the uncertainties of the parameters. It is important to remember that in the real world we would only see B and C since the true behavior (A) is masked by measurement error.

neurobiological model's arrows in Figure 1B thus is more appropriately represented not by a linear regression but rather by a differential equation, which can express how the activation of each node changes in time as a function of the activations of all nodes that interact with it (as per the control circuit diagram shown in Figure 1C). In such a case, the model is comprised of all the linked ("coupled") differential equations and their parameters, and the solution of such a differential equation given the initial conditions (typically all activations at time=0) is completely determined by the parameters. The task of model fitting is thus to find the best parameters given noisy data. While for linear regression the "model" is determined by least squares fit in which we attempt to minimize the distance ("noise") between the regression ("model") and the data points, in a dynamical system "noise" is quantified somewhat differently. Figure 3 illustrates the procedure of model fitting and validation for a simple deterministic dynamical system that describes a typical interplay between excitatory and inhibitory components of a simple circuit with negative feedback. Figure 3A reflects the "true" state of the system, but due to measurement error we can only obtain Figure 3B. From the data in Figure 3B we can estimate the initial conditions and parameters, but because of noise both of these carry uncertainties. Figure 3C shows a sample of possible solutions given the fitted initial conditions and parameters. That means that we can validate our model by confirming that the validation experiments fall within the set of possible solutions.

In dynamical systems, not only measurement noise but also intrinsic noise (reflecting the probabilistic nature of biological processes) turns our differential equations into stochastic differential equations without losing the ability to predict, and therefore to validate, models. Stochastic differential equations are differential equations with noise terms that purposefully introduce randomness into the system. This means that even if we know the exact initial conditions and all the parameters, each solution will be slightly different. Model fitting in this case is illustrated in Figure 4 using a simple stochastic model that describes neural activation fluctuations around a homeostatic steady state value. Figure 4A shows a measurement that reflects a certain set of parameters and initial conditions. From these data we can fit the most likely initial conditions and parameters, which again are uncertain. Figure 4B shows 5 equivalent solutions for which we drew initial conditions and parameters from their respective distributions. We validate the accuracy of our model by determining if new experiments produce trajectories that fall within the set of possible solutions provided by our initial fit (Figure 4C). To objectively and quantitatively assess the model's utility, its power is determined by the time duration over which actual trajectories continue to fall within the set of predicted ones as well as by the degree to which that consistency continues to exist over a diversity of new inputs.

Importantly, this process of model assessment in the face of noise follows a logic that is Bayesian, rather than the "frequentist" statistical approach more common to psychiatric models. Rather than asking the question "What is the probability that our data resulted from the null hypothesis?" (as per P value statistics) Bayesian analyses ask, "Given the data, what are the values and uncertainties of my model parameters?" (Sivia and Skilling, 2006). The corresponding probability distribution of model parameters is the posterior distribution. The standard procedure to obtain the posterior distributions is to express the likelihood of the data given the parameters of the model. By maximizing the likelihood with respect to the parameters, we can then identify the most likely parameters of the model and their corresponding uncertainties.

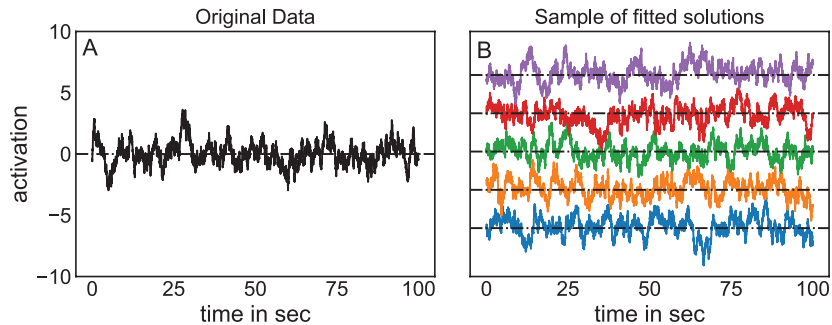

**Figure 4.** Example of how to model/build a model with intrinsic stochasticity. Here we present a stochastic dynamical system that could be used to model the brain activity of a single node in the resting-state-network (Ornstein-Uhlenbeck [OU] process) (Strey, 2019). (A) A particular solution of the OU process with amplitude = 1 and relaxation time = 1 second. (B) From the data on the left, we estimated the model parameters of the OU process and created a set of sample solutions. Because of the stochastic nature of the OU process, none of the solutions look identical even though their parameters are the same. The process of model validation determines how likely it is that the data of a validation experiment could have originated from the fitted parameters, an assessment that—due to its logical structure—must be determined using Bayesian, rather than frequentist, statistical tests.

While this approach may appear to be only semantically different than standard (frequentist) ways of analyzing data, in fact the mathematical and conceptual differences are profound, for 3 reasons. First, unlike standard statistics, it permits accurate estimation of models from timeseries (Strey, 2019). While correlations between time-series today are ubiquitous within neuroimaging analyses (indeed, they form the very basis of resting-state analyses, including all "network" results), correlations between time-series are inappropriate given that correlations assume the independence of data points, and data points within a time-series are, by definition, not independent—an error that artificially inflates the statistical significance of *P* values. A second advantage is that posterior probabilities can be used to compare competing models by testing which model is more likely to be correct given the available data (Sivia and Skilling, 2006). Third, a Bayesian approach can also help us in assessing the validity of model systems, including nonlinear systems when, as described earlier in the context of the billiard balls or circuits, we have a long chain of interactions $x \Rightarrow y \Rightarrow z$ … or feedback loops. For such a configuration, we would formulate a model that contains all of the interaction parameters between the nodes. But instead of only considering pairwise interactions, we maximize the likelihood of the complete set of activations to find the optimal set of parameters. This means that activations further down the chain constrain and improve the estimates of parameters of earlier interactions, thereby preventing the accumulation of errors along the chain.

## Building Computational Psychiatry Models

In 1609, Johannes Kepler published the first and second law of planetary motions, which stated that planets travel on ellipses and that the area speed of planets is constant. He derived this law by analyzing observations of planetary movements by Tycho Brahe, noting that all observed planets shared these particular properties. Later, Newton, in his *Principia* published in 1687, recognized that Kepler's laws of planetary motion can all be explained by the radial nature and distance dependence of the gravitational force between the planets and the sun. Here, Newton reconciled his theory with existing data and analysis by Kepler but simplified the model to only requiring a single gravitational force that is consistent with all observations. In that sense, Newton's model was more powerful because it could not only explain, but also predict, more data with fewer parameters.

Unfortunately, there are no generally accepted rules on how to build good models. On the other hand, there do exist well-established reverse-engineering strategies that are used in systems biology (Csete and Doyle, 2002; Tomlin and Axelrod, 2005) and can act as a guide. In recent years, tremendous progress has been made in understanding control circuits in cells (both prokaryotic and eukaryotic), a success grounded in the application of control systems engineering principles to biology. Cells, in order to survive, need to maintain homeostasis in the face of a chaotic environment, which they accomplish through control circuits (negative feedback loops with excitatory and inhibitory control), structures whose modeling is well understood from control systems engineering. The first success for this approach of searching for closed-loop biological control circuits came from bacterial chemotaxis (Alon et al., 1999). El-Samad et al. in 2005 identified several feedback systems in the heat shock response in cells, and the researchers were able by simulations and experiments to decompose the circuit into intuitively comprehensible subsystems (El-Samad et al., 2005). In 2018, Zhang et al. developed a fully dynamic system for the cell cycle that helped identify the key elements for controlling the cell-cycle speed (Zhang and Wang, 2018).

Several recent reviews on computational psychiatry (Wang and Krystal, 2014; Huys et al., 2016) have focused on how computational methods, using both machine learning and model-based approaches, could be used to diagnose mental diseases but also predict treatment strategies. As one review (Huys et al., 2016) has pointed out, machine learning applied by itself is not a very useful tool for these tasks since the data, typically M/EEG or fMRI time series, have high dimensionality (i.e., the data extend over many points in time and space). These data tend to overfit the model, which reduces the model's ability to predict. The authors suggest that by using known theoretical models of the underlying disorder, one could reduce the dimensionality of the data (e.g., by reducing the number of brain regions that are contributing to the disorder, or by reducing the dynamics of the data into characteristic parameters) and therefore improve the machine learning predictability. This method has been successfully applied to clustering schizophrenia patients into subgroups using a working memory task while undergoing fMRI (Brodersen et al., 2014).

However, if we are correct in our assumption that psychiatry is a system about which we still know very little, then we should be wary of committing category errors in constraining our data-driven machine learning. In most cases, the very brain regions and/or the structure of the network responsible for a particular psychiatric disease have not been unequivocally validated, making them risky assumptions for further model-building. One strategy for addressing this risk, with a minimum of assumptions, is to pursue reverse-engineering strategies found in the emerging field of "scientific machine learning," which seeks to identify physical and biological laws from data in an automated fashion (Daniels and Nemenman, 2015; Daniels et al., 2019). In particular, several methods have recently been developed to use high-dimensional dynamic (time-series) data to find underlying patterns. Some of the most promising techniques for potential applications in brain research are dynamic mode decomposition (Schmid, 2010; Taira et al., 2017), which identifies spatial-dynamical pattern in an automated fashion, and sparse identification of nonlinear dynamics (Brunton et al., 2016), which allows the identification of nonlinear differential equations that are compatible with the data. The challenge arises when we wish to use data-driven approaches to tell us something conceptually meaningful about the biology. Variational autoencoders are a type of generative machine learning that can be used to identify a latent structure underlying complex data (the implicit "rules" that govern such data, which can then be used to "generate" predictions, and therefore validation experiments, from those rules). However, it is quite likely that these latent structures are purely mathematical entities that do not map onto corresponding biological structures.

Here, a hybrid approach for circuit discovery may be helpful. Conceptually, circuit discovery is grounded in the observation that a circuit's output dynamics map onto a differential equation whose mathematical form can itself provide purely design clues as to the system's mechanistic structure (**Figure 5**). Thus, using inputs $u$ and outputs $y$, one can use system identification (Ljung, 2014) to generate the differential equation that describes the system's response, known as its transfer function. Once the transfer function has been established, machine learning then can be used to pattern-match the transfer function onto a finite set of biologically plausible circuit motifs (e.g., series [**Figure 5A**], parallel [**Figure 5B**], and feedback [**Figure 5C**] motifs) that generate equivalent dynamics.

After defining inputs and outputs, the next step is to refine the model to include 3-node interactions as required for closed-loop circuits. To do so, we need to identify contributions of indirect interactions between 2 brain regions that are caused by a third region interacting with both nodes. To illustrate, let us say that we have brain regions A and B and can measure the activation in both. However, there may be a third region, C, that interacts with both A and B; if so, how much of the A-B interaction is due to C? Since all system interactions can be written mathematically as differential equations, we can check at every step not only what the most likely parameters of our model are, but also how well our model describes the data. This process of expanding the function and structure to include more and more regions will continue until the model includes all the contributing nodes. Unlike our statistical example, in which every node with one or more inputs must be described by a separate model (i.e., a multiple linear regression), our system of coupled differential equations describes the system as a whole, including all of its relationships (structure) and interactions (function). The fact that we have a mathematical description of the complete system thus permits us to validate the model, by simulating the temporal response of the network with arbitrary inputs, to make predictions that can then be tested against independent data. In doing so, we can incorporate measurement and intrinsic noise sources, including inter-subject variance, within the model by using model validation and Bayesian model selection to adapt the model to a system of coupled stochastic (i.e., probabilistic) differential equations.

## Computational Psychiatry: Constructing Circuit-Based Features for Research Domain Criteria (RDoC)?

In 2010, the National Institute of Mental Health launched the initiation of a new psychiatric "grammar," based on objective, observable, and measurable (e.g., behavioral and biological) features, arranged in taxonomies of scale and function. These research domain criteria (RDoC) were motivated by the need to accelerate integration of disparate research approaches by permitting use of a shared set of features relevant to understanding both normal mental processes as well as their disorders. RDoC was based on the explicit assumption that "mental disorders are biological disorders involving brain circuits that implicate specific domains of cognition, emotion, or behavior" (https://www.nimh.nih.gov/about/directors/thomas-insel/blog/2013/transforming-diagnosis.shtml). As such, circuit regulation may constitute one such feature.

One striking aspect of psychiatric neuroimaging results is the fact that they overwhelmingly implicate the same 3 circuits (which partially intersect and are mutually interacting) for nearly all psychiatric disorders: those that involve processing of threat (e.g., the amygdala, hippocampus, orbitofrontal cortex, ventromedial prefrontal cortex), reward (e.g., the amygdala, ventral tegmental area, locus coeruleus, nucleus accumbens), and perceptual stimuli (e.g., the thalamus, sensory cortex and inferior frontal gyrus). This tendency to implicate the same regions of interest across multiple psychiatric diagnoses is one clue that the distinct flavors of mental disorders are probably not rooted simply in a region or its connectivity being stronger or weaker but rather are likely to implicate distinct modes of deregulatory control processes between them—that is, modeled by dynamical systems. Importantly, these control processes may reflect disruptions in neurotransmitters and/or other molecular mechanisms and yet be effectively dynamically modeled as emergent phenomena at the macroscopic scale. In dynamical systems, it is common to represent the full set of possible physical states within an n-dimensional phase space. If the

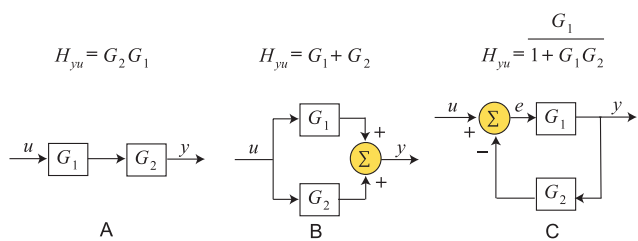

$$H_{yu} = G_2 G_1 \qquad H_{yu} = G_1 + G_2 \qquad H_{yu} = \frac{G_1}{1 + G_1 G_2}$$

**Figure 5.** Circuit discovery. To illustrate how transfer function structure changes with different circuit topologies, we show 3 transfer functions, each of which corresponds to a different kind of "motif," with series (A), parallel (B), and feedback (C) circuits. By using pairs of inputs (u) and outputs (y) to obtain their transfer function (system identification), we can systematically infer circuit topology. A hybrid version uses machine learning to pattern-match data dynamics to the canonical dynamics produced by each motif. This type of data-driven reverse engineering has been used successfully across science and engineering domains, including systems biology (Bongard and Lipson, 2007; Luo et al., 2010; Brunton et al., 2016).

full spectrum of regulation is represented along an axis designated for each circuit, then—in principle—we can position each psychiatric disorder at precise coordinates within such a phase space. For example, we have shown that the anxious-to-reckless spectrum tracks feedback dynamics within the threat-detection and perceptual circuits (Mujica-Parodi et al., 2017). In contrast, depression likely implicates feedback dynamics with the reward circuit, and complex disorders such as paranoid schizophrenia and addiction are likely to implicate all 3 circuits and potentially others as well. Computational psychiatry demonstrates that the same circuit can be dysregulated in multiple ways, leading to very different clinical outcomes. Thus, computational psychiatry can expand RDoC's grammar to move beyond features to include circuits and the dynamical motifs that represent various ways in which their signal processing can be disrupted.

## Conclusions

By using systems of coupled (stochastic) differential equations, rather than linear regressions, computational psychiatry is able to model the complex causal chains and feedback loops thought to underlie neuropsychiatric circuits. Thus, computational psychiatry extends the reach of current psychiatric conceptual frameworks not just by applying machine learning towards better discrimination of diagnostic categories, but more profoundly by providing models that, because they are predictive, now can be rigorously assessed for their accuracy and power. Doing so would extend psychiatry's intuition that disorders result from emergent brain circuits, implicit in the neuroimaging literature, beyond purely conceptual schemas to align with fundamental physiological principles of homeostatic regulation and feedback found in other areas of "dynamic disease" (Mackey and Glass, 1977). The benefit of this control systems approach is that it is sufficiently biomimetic to interface directly with clinical neuroimaging, providing a path towards reconciliation of mainstream concepts of "circuits" and "regulation" with more mechanistic processes testable with animal probes. But the most potentially transformative aspect of computational psychiatry is clinical. Dynamical systems are designed to predict future trajectories and thus offer the tantalizing possibility of models that may one day accurately predict the timing of a first psychiatric break, the frequency of bipolar cycling, or the response lag between Drug A versus Drug B. Thus, computational psychiatry can provide for psychiatry a version of what Newtonian mechanics provided for physics: in this case, robust and useful models for how the brain transitions over time between states of health and disease.

## Acknowledgments

This work was supported by the National Institute of Drug Abuse (1 R44 DA043277-01), the W.M. Keck Foundation, and the National Science Foundation **Brain Research through Advancing Innovative Neurotechnologies (BRAIN)** Initiative (1926781).

## Statement of Interest

None.

## References

Alon U (2019) An introduction to systems biology: design principles of biological circuits. 2nd ed. Boca Raton, FL: Chapman and Hall/CRC.

Alon U, Surette MG, Barkai N, Leibler S (1999) Robustness in bacterial chemotaxis. Nature 397:168–171.

Bartz J, Simeon D, Hamilton H, Kim S, Crystal S, Braun A, Vicens V, Hollander E (2011) Oxytocin can hinder trust and cooperation in borderline personality disorder. Soc Cogn Affect Neurosci 6:556–563.

Bassett DS, Greenfield DL, Meyer-Lindenberg A, Weinberger DR, Moore SW, Bullmore ED (2010) Efficient physical embedding of topologically complex information processing networks in brains and computer circuits. Plos Comput Biol 6:e1000748.

Baumgartner T, Heinrichs M, Vonlanthen A, Fischbacher U, Fehr E (2008) Oxytocin shapes the neural circuitry of trust and trust adaptation in humans. Neuron 58:639–650.

Benoit D (2004) Infant-parent attachment: Definition, types, antecedents, measurement and outcome. Paediatr Child Health 9:541–545.

Bongard J, Lipson H (2007) Automated reverse engineering of nonlinear dynamical systems. Proc Natl Acad Sci U S A 104:9943–9948.

Breakspear M (2017) Dynamic models of large-scale brain activity. Nat Neurosci 20:340–352.

Brodersen KH, Deserno L, Schlagenhauf F, Lin Z, Penny WD, Buhmann JM, Stephan KE (2014) Dissecting psychiatric spectrum disorders by generative embedding. Neuroimage Clin 4:98–111.

Brunton SL, Proctor JL, Kutz JN (2016) Discovering governing equations from data by sparse identification of nonlinear dynamical systems. Proc Natl Acad Sci U S A 113:3932–3937.

Bullmore E, Sporns O (2009) Complex brain networks: graph theoretical analysis of structural and functional systems. Nat Rev Neurosci 10:186–198.

Csete ME, Doyle JC (2002) Reverse engineering of biological complexity. Science 295:1664–1669.

Daniels BC, Nemenman I (2015) Automated adaptive inference of phenomenological dynamical models. Nat Commun 6:8133.

Daniels BC, Ryu WS, Nemenman I (2019) Automated, predictive, and interpretable inference of Caenorhabditis elegans escape dynamics. Proc Natl Acad Sci U S A 116:7226–7231.

Deco G, Jirsa VK, McIntosh AR (2011) Emerging concepts for the dynamical organization of resting-state activity in the brain. Nat Rev Neurosci 12:43–56.

De Dreu CK, Greer LL, Handgraaf MJ, Shalvi S, Van Kleef GA, Baas M, Ten Velden FS, Van Dijk E, Feith SW (2010) The neuropeptide oxytocin regulates parochial altruism in intergroup conflict among humans. Science 328:1408–1411.

De Dreu CK, Greer LL, Van Kleef GA, Shalvi S, Handgraaf MJ (2011) Oxytocin promotes human ethnocentrism. Proc Natl Acad Sci U S A 108:1262–1266.

Domes G, Heinrichs M, Gläscher J, Büchel C, Braus DF, Herpertz SC (2007) Oxytocin attenuates amygdala responses to emotional faces regardless of valence. Biol Psychiatry 62:1187–1190.

El-Samad H, Kurata H, Doyle JC, Gross CA, Khammash M (2005) Surviving heat shock: control strategies for robustness and performance. Proc Natl Acad Sci U S A 102:2736–2741.

Frank MJ (2005) Dynamic dopamine modulation in the basal ganglia: a neurocomputational account of cognitive deficits in medicated and nonmedicated Parkinsonism. J Cogn Neurosci 17:51–72.

Gamer M, Zurowski B, Büchel C (2010) Different amygdala subregions mediate valence-related and attentional effects of oxytocin in humans. Proc Natl Acad Sci U S A 107:9400–9405.

Grillon C, Krimsky M, Charney DR, Vytal K, Ernst M, Cornwell B (2013) Oxytocin increases anxiety to unpredictable threat. Mol Psychiatry 18:958–960.

Gu S, Pasqualetti F, Cieslak M, Telesford QK, Yu AB, Kahn AE, Medaglia JD, Vettel JM, Miller MB, Grafton ST, Bassett DS (2015) Controllability of structural brain networks. Nat Commun 6:8414.

Huys QJ, Maia TV, Frank MJ (2016) Computational psychiatry as a bridge from neuroscience to clinical applications. Nat Neurosci 19:404–413.

Ide JS, Nedic S, Wong KF, Strey SL, Lawson EA, Dickerson BC, Wald LL, La Camera G, Mujica-Parodi LR (2018) Oxytocin attenuates trust as a subset of more general reinforcement learning, with altered reward circuit functional connectivity in males. Neuroimage 174:35–43.

Kirsch P, Esslinger C, Chen Q, Mier D, Lis S, Siddhanti S, Gruppe H, Mattay VS, Gallhofer B, Meyer-Lindenberg A (2005) Oxytocin modulates neural circuitry for social cognition and fear in humans. J Neurosci 25:11489–11493.

Kosfeld M, Heinrichs M, Zak PJ, Fischbacher U, Fehr E (2005) Oxytocin increases trust in humans. Nature 435:673–676.

Ljung L (2014) System identification toolbox: user's guide. Natick, MA: Mathworks.

Luo J, Wang J, Ma TM, Sun Z (2010) Reverse engineering of bacterial chemotaxis pathway via frequency domain analysis. Plos One 5:e9182.

Mackey MC, Glass L (1977) Oscillation and chaos in physiological control systems. Science 197:287–289.

Maia TV, Frank MJ (2011) From reinforcement learning models to psychiatric and neurological disorders. Nat Neurosci 14:154–162.

Manita S, Suzuki T, Homma C, Matsumoto T, Odagawa M, Yamada K, Ota K, Matsubara C, Inutsuka A, Sato M, Ohkura M, Yamanaka A, Yanagawa Y, Nakai J, Hayashi Y, Larkum ME, Murayama M (2015) A top-down cortical circuit for accurate sensory perception. Neuron 86:1304–1316.

Mease RA, Krieger P, Groh A (2014) Cortical control of adaptation and sensory relay mode in the thalamus. Proc Natl Acad Sci U S A 111:6798–6803.

Morris JA (2012) The fall of the schizophrenogenic mother. Lancet 380:110.

Mujica-Parodi LR, Cha J, Gao J (2017) From anxious to reckless: a control systems approach unifies prefrontal-limbic regulation across the spectrum of threat detection. Front Syst Neurosci 11:18.

Murphy KF, Adams RM, Wang X, Balázsi G, Collins JJ (2010) Tuning and controlling gene expression noise in synthetic gene networks. Nucleic Acids Res 38:2712–2726.

Murray JD, Demirtaş M, Anticevic A (2018) Biophysical modeling of large-scale brain dynamics and applications for computational psychiatry. Biol Psychiatry Cogn Neurosci Neuroimaging 3:777–787.

Ne'eman R, Perach-Barzilay N, Fischer-Shofty M, Atias A, Shamay-Tsoory SG (2016) Intranasal administration of oxytocin increases human aggressive behavior. Horm Behav 80:125–131.

Petrovic P, Kalisch R, Singer T, Dolan RJ (2008) Oxytocin attenuates affective evaluations of conditioned faces and amygdala activity. J Neurosci 28:6607–6615.

Pillai AS, Jirsa VK (2017) Symmetry breaking in space-time hierarchies shapes brain dynamics and behavior. Neuron 94:1010–1026.

Rădulescu A, Mujica-Parodi LR (2014) Network connectivity modulates power spectrum scale invariance. Neuroimage 90:436–448.

Schmid PJ (2010) Dynamic mode decomposition of numerical and experimental data. J Fluid Mech 656:5–28.

Schurmann F, Hill S, Markram H (2007) The Blue Brain Project: building the neocortical column. BMC Neuroscience 8:P109.

Shamay-Tsoory SG, Fischer M, Dvash J, Harari H, Perach-Bloom N, Levkovitz Y (2009) Intranasal administration of oxytocin increases envy and schadenfreude (gloating). Biol Psychiatry 66:864–870.

Sivia D, Skilling J (2006) Data analysis: a Bayesian tutorial. 2nd ed. New York, NY: Oxford University Press.

Strey HH (2019) Estimation of parameters from time traces originating from an Ornstein-Uhlenbeck process. Phys Rev E 100:062142.

Taira K, Brunton SL, Dawson STM, Rowley CW, Colonius T, McKeon BJ, Schmidt OT, Gordeyev S, Theofilis V, Ukeiley LS (2017) Modal analysis of fluid flows: an overview. AIAA J 55:4013–4041.

Tomlin CJ, Axelrod JD (2005) Understanding biology by reverse engineering the control. Proc Natl Acad Sci U S A 102:4219–4220.

Wang XJ, Krystal JH (2014) Computational psychiatry. Neuron 84:638–654.

Zhang K, Wang J (2018) Exploring the underlying mechanisms of the xenopus laevis embryonic cell cycle. J Phys Chem B 122:5487–5499.